\documentclass[twocolumn]{aastex631}

\newcolumntype{R}{>{$}r<{$}}

\newcommand{\newsnr}{\hat{\rho}}
\newcommand{\Msun}{M_\odot}

\begin{document}
\title[PyCBC Live for O3]{Realtime search for compact binary mergers in Advanced LIGO and Virgo's third observing run using PyCBC Live}

\author[0000-0001-5078-9044]{Tito Dal Canton}
\affiliation{NASA Goddard Space Flight Center, 8800 Greenbelt Rd., Greenbelt, MD 20771, USA}
\affiliation{Max Planck Institute for Gravitational Physics (Albert Einstein Institute),
  Am M\"uhlenberg 1, D-14476 Potsdam-Golm, Germany}
\affiliation{Universit\'e Paris-Saclay, CNRS/IN2P3, IJCLab, 91405 Orsay, France}

\author[0000-0002-1850-4587]{Alexander H. Nitz}
\affiliation{Max Planck Institute for Gravitational Physics (Albert Einstein Institute), D-30167 Hannover, Germany}
\affiliation{Leibniz Universit\"at Hannover, D-30167 Hannover, Germany}

\author[0000-0002-1534-9761]{Bhooshan Gadre}
\affiliation{Max Planck Institute for Gravitational Physics (Albert Einstein Institute),
  Am M\"uhlenberg 1, D-14476 Potsdam-Golm, Germany}

\author[0000-0002-4289-3439]{Gareth S. Cabourn Davies}
\affiliation{Instituto Galego de F\'{i}sica de Altas Enerx\'{i}as, Universidade de Santiago de Compostela, 15782 Santiago de Compostela, Galicia, Spain}
\affiliation{Institute for Cosmology and Gravitation, University of Portsmouth, 1-8 Burnaby Road,
         Portsmouth, P01 3FZ, UK}

\author[0000-0001-7983-1963]{Ver\'{o}nica Villa-Ortega}
\affiliation{Instituto Galego de F\'{i}sica de Altas Enerx\'{i}as, Universidade de Santiago de Compostela, 15782 Santiago de Compostela, Galicia, Spain}

\author[0000-0003-1354-7809]{Thomas Dent}
\affiliation{Instituto Galego de F\'{i}sica de Altas Enerx\'{i}as, Universidade de Santiago de Compostela, 15782 Santiago de Compostela, Galicia, Spain}

\author[0000-0002-5304-9372]{Ian Harry}
\affiliation{Max Planck Institute for Gravitational Physics (Albert Einstein Institute),
  Am M\"uhlenberg 1, D-14476 Potsdam-Golm, Germany}
\affiliation{Institute for Cosmology and Gravitation, University of Portsmouth, 1-8 Burnaby Road,
         Portsmouth, P01 3FZ, UK}

\author[0000-0003-2703-449X]{Liting Xiao}
\affiliation{LIGO Laboratory, California Institute of Technology, MS 100-36, Pasadena, California 91125, USA}

\begin{abstract}
The third observing run of Advanced LIGO and Advanced Virgo took place between
April 2019 and March 2020 and resulted in dozens of gravitational-wave candidates,
many of which are now published as confident detections.
A crucial requirement of the third observing run has been the rapid
identification and public reporting of compact binary mergers, which enabled
massive followup observation campaigns with electromagnetic and neutrino
observatories.
PyCBC Live is a low-latency search for compact binary mergers based on
frequency-domain matched filtering, which has been used during the second and
third observing runs, together with other low-latency analyses, to generate
these rapid alerts from the data acquired by LIGO and Virgo.
This paper describes and evaluates the improvements made to PyCBC Live after the
second observing run, which defined its operation and performance during the
third observing run.
\end{abstract}

\section{Introduction}
\label{sec:intro}

The Advanced LIGO and Advanced Virgo gravitational-wave (GW) observatories
conducted their first two observing runs, O1 and O2, between 2015 and 2017
\citep{AdvLIGO,AdvVirgo,ObsScenario}.
From these two runs, over a dozen confident observations of binary black hole
(BBH) mergers and one binary neutron star (BNS) merger were made
\citep{TheLIGOScientific:2017qsa, GWTC1, Nitz:2018imz, Nitz:2019hdf,
Venumadhav:2019tad, Zackay:2019btq, Venumadhav:2019lyq, Nitz:2020naa}.
The BNS merger, GW170817, was observed within a minute of the data being
recorded, was associated with a gamma ray burst \citep{Monitor:2017mdv} and was
subsequently followed up by a large number of observatories spanning the whole
electromagnetic (EM) spectrum \citep{GBM:2017lvd}.
Without the realtime identification and localization of GW170817 as a merging
pair of neutron stars, these observations would not have been possible.

The third observing run of Advanced LIGO and Advanced Virgo, O3, began in
April 2019 and ended in March 2020.
Thanks to hardware improvements in the detectors, the ranges for compact binary
mergers were expected to increase by 6--85\% with respect to O2, depending on
the source type and detector, with the largest improvement expected for Virgo
\citep{ObsScenario}.
The sensitive volume of the run was then predicted to be $3.3 \times 10^6$
Mpc$^3$ yr for BNS mergers, and $3.4 \times 10^8$ Mpc$^3$ yr for BBH mergers.
With this unprecedented sensitivity, we expected to observe up to $\approx 40$
BBH mergers, up to $\approx 10$ BNS mergers, and potentially make the first
observation of a neutron-star-black-hole (NSBH) merger.
Indeed, dozens of candidates from O3 have been uploaded to
GraceDB\footnote{\href{https://gracedb.ligo.org/superevents/public/O3}{https://gracedb.ligo.org/superevents/public/O3}}
and announced on the Gamma-ray Coordinates
Network\footnote{\href{https://gcn.gsfc.nasa.gov}{https://gcn.gsfc.nasa.gov}} (GCN).
Four such candidates have been published as notable compact binary mergers
so far \citep{GW190425,GW190412,GW190814,GW190521}.
Many more confident detections from the first half of O3 have also been
presented in the GWTC-2 \citep{GWTC2} and 3-OGC \citep{3OGC} catalogs.
O3 was the first observing run in which three observatories operated for
the full duration of the run, increasing the observing duty cycle of the
network, reducing the uncertainty in the sky location of observed events and
therefore maximizing the chance of making a multi-messenger
observation \citep{ObsScenario}.
Rapid processing of data from all three observatories was therefore a crucial
requirement.

PyCBC Live, first introduced by \cite{Nitz:2018rgo}, is one of several search
codes analysing the data in realtime to observe compact binary mergers, other
analyses being GstLAL \citep{Messick:2016aqy,Sachdev:2019vvd}, MBTA
\citep{Adams:2015ulm,Aubin:2020goo}, SPIIR
\citep{Hooper:2011rb,Chu2017,Chu:2020pjv} and the generic-transient search cWB
\citep{PhysRevD.93.042004}.
Having multiple analyses provides for redundancy, both in terms of the
possibility that one of the analysis fails for any reason, and in terms of the
independent methodology that each of these analyses applies to identify compact
binary mergers.
PyCBC Live is based on the more general PyCBC software package
\citep{PyCBCZenodo} and uses a precalculated bank of compact binary merger
waveform models combined with matched filtering in the frequency domain
\citep{FindChirp, Babak:2012zx}.
PyCBC Live has been instrumental in many of the GW observations to date, both in
O2 \citep{GW170104,GW170608,GW170814,TheLIGOScientific:2017qsa} and O3
\citep{GW190412, GW190425, GW190521}.

In this paper we describe the improvements that have been made to PyCBC Live in
preparation for, and during, O3.
Specifically, we discuss the new techniques that (i) enabled the simultaneous
and symmetric analysis of data from three observatories, and the reliable
assessment of the statistical significance of observed candidates, (ii)
improvements in the handling of instrumental transients, (iii) an updated
technique to detect signals in data from a single detector and (iv) a method to
rapidly infer the nature of the compact objects involved in a candidate merger.
We then evaluate the effect of these improvements in terms of search
sensitivity by simulating compact binary signals in Gaussian data at the design
sensitivity of advanced GW detectors, as well as in real O3 data from Advanced
LIGO and Advanced Virgo. We also evaluate the accuracy of the source
classification method and the effect of these improvements on the latency of
the produced candidates using O2 data.

\section{New methods for the third observing run}
\label{sec:methods}

This section describes the new methodology that was used or developed in
PyCBC Live during O3. Each subsection describes changes to a specific aspect of
the analysis, and subsections are ordered so as to follow the data flow through
the pipeline as much as possible.

\subsection{Search space and template bank}
\label{ssec:bank}

Although it is an improvement to the configuration of PyCBC Live rather
than the code itself, we begin by describing the search space and template bank
adopted during O3, as a reference for future work.

The bank covers the same mass, spin and waveform duration space as that
proposed by \cite{DalCanton:2017ala} and previously adopted during O2.
The same waveform models are also employed: TaylorF2 \citep{Buonanno:2009zt,
Bohe:2013cla} for BNS and low-mass NSBH templates, and a reduced-order
frequency-domain version of SEOBNRv4 \citep{PhysRevD.95.044028, Purrer:2014fza}
for BBH and heavy NSBH templates.

However, the O3 bank utilises a template placement method based on an
optimized hybrid geometric-random approach \citep{Roy:2017qgg, Roy:2017oul,
HybridBankCOSPAR} which is more efficient than the ``manual'' combination of
geometric and stochastic methods used for the O2 bank \citep{DalCanton:2017ala}
in the sense that the obtained bank is $\sim 25$\% smaller and can be produced
much faster.
As a result, the O3 bank is only 13\% larger than the O2 bank, despite the
increased sensitivity of the detectors.

\subsection{Improved rejection of instrumental transients}
\label{ssec:glitches}

Loud instrumental transients in GW data (\emph{glitches}) which last much less
than a second can corrupt the results of transient searches on a time scale much
longer than the glitch itself \citep{DalCanton:2013joa}.
In particular, due to the relatively long-lasting impulse response of the
various filters involved in the SNR calculation, loud glitches can cause the SNR
time series for a given template to cross the trigger-generation threshold many
times over several seconds. The resulting high-SNR triggers then dominate over
quieter triggers from the underlying stationary noise and possible
astrophysical signals, effectively blinding the search for the entire duration
of the impulse response of the filters. In early O3, this issue appeared
prominently in the results of PyCBC Live as occasional gaps in the production
of triggers from a given detector, lasting from a few seconds to several tens
of seconds, depending on the glitch.

A simple and widely-used solution to this
problem, called \emph{gating}, consists of windowing out the GW strain data for a
short window centered on the glitch prior to matched filtering
\citep{TheLIGOScientific:2016qqj,Usman:2015kfa,Sachdev:2019vvd}.  PyCBC offline
searches already implement this method by detecting glitches as loud excursions
in the whitened strain, and then multiplying the data with the complement of a
Tukey window centered on the glitch time.  We adopted the same algorithm in
PyCBC Live during O3. We used a threshold of $50\sigma$ on the absolute value
of the whitened strain time series as a glitch detector. Each detected glitch
was gated with a symmetric complemented Tukey window, configured to have 0.5 s
of central zeroes and 0.25 s of smooth taper on both sides. This approach
significantly reduced the impact of high-SNR non-Gaussian transient noise with
no visible impact on the latency of the analysis.
The chosen gating duration is justified because many high-SNR glitches
tend to be shorter than 1 s \citep{LIGO:2021ppb} and significantly longer gates
might cause more damage to downstream stages of the analysis than the glitch
itself.
A fixed duration also removes the need to estimate the duration of the glitch,
which is nontrivial due to the impulse response of the whitening filter, often
longer than the glitch itself.
Nevertheless, a fast gating procedure which more accurately identifies the
time-frequency extent of a glitch could be beneficial in the future.

Another improvement inherited from PyCBC's offline search during O3 was the
inclusion of the high-frequency sine-Gaussian discriminator in
the ranking of single-detector triggers.  The discriminator, described in
\citet{Nitz:2017lco}, exploits the fact that most compact binary mergers
induce a negligible amount of signal power at frequencies higher than
the ringdown of the dominant quadrupole mode.  By measuring the excess power at
the time of peak signal amplitude, and at various frequencies above the
ringdown, a $\chi^2$ statistic is constructed and used to reweight the
single-detector trigger ranking statistic. The discriminator is most effective at
preventing glitches from triggering high-mass templates with final frequencies
of $\approx 100$\,Hz or less, hence it increases the search sensitivity to
high-mass black hole mergers.
We adopted the same implementation of the discriminator used by PyCBC's offline search
with a negligible impact on PyCBC Live's latency.

\subsection{Inclusion of Virgo in the coincident search}
\label{ssec:methods_multi}

Advanced Virgo began operating in conjunction with the LIGO observatories in
the last few months of O2, when its sensitivity was typically $1/4$ to $1/3$
that of the LIGO instruments.
Despite the smaller sensitivity, the inclusion of Virgo markedly improved the
localization of important candidates, such as GW170817 \citep{TheLIGOScientific:2017qsa}.
However, Virgo's contribution to the overall network sensitivity was limited,
as quantified by its integrated BNS observed time-volume of $4 \times 10^3$\,Mpc$^3$\,yr
compared to $5 \times 10^5$\,Mpc$^3$ yr for LIGO-Hanford\footnote{See
\href{https://www.virgo-gw.eu/O2.html}{https://www.virgo-gw.eu/O2.html} and
\href{https://www.gw-openscience.org/detector_status/O2}{https://www.gw-openscience.org/detector\_status/O2}.}.
Hence, in order to produce candidate events, the PyCBC Live analysis
introduced in \cite{Nitz:2018rgo} relied on a simple coincident detection
between the two LIGO observatories only.
Additional detectors were analyzed by PyCBC Live for the purpose of improving
the rapid spatial localization, but they did not contribute to the significance
of candidates, and they could not produce candidates in coincidence with one of
the LIGO detectors.

However, as the relative sensitivities of the instruments within the global
GW network become comparable, each instrument's contribution to the overall
sensitivity also increases.
In the coming years, the Virgo observatory may approach $60$--$80\%$ of the
LIGO detector sensitivities, limited primarily by its shorter arm length \citep{ObsScenario}.
In addition, new instruments will be joining the global network: KAGRA
\citep{Somiya:2011np,Aso:2013eba,Akutsu:2020his}, which conducted its first
observing period shortly after the end of O3, and LIGO-India \citep{LIGOIndia},
scheduled to begin operation in the mid 2020's \citep{ObsScenario}.
The higher network sensitivity comes about from two sources: (1) improvement
in overall network uptime due to overlap between instruments live time, and
(2) improvement in detection confidence (reduction of the false-alarm rate)
from additional detectors.
Hence, in order to start exploiting the benefits of a larger and more
symmetric network, the PyCBC Live analysis has been modified for the O3 run.

In its O3 configuration, PyCBC Live correlates the full bank of template
waveforms with data from all operating detectors.
Then for each detector pair, we independently perform the same double-coincident
analysis used for LIGO-Hanford and LIGO-Livingston during O2.\footnote{Note that
the ranking of double-coincident triggers in each detector pair depends on the
pair, since it involves the distribution of expected relative signal phases,
times and amplitudes appropriate for the two detectors \citep{Nitz:2017svb}.}
When a pair of detectors labelled $A$ and $B$ observe a coincident candidate, a
false-alarm rate $\mathcal{F}_{AB}$ is computed using time-shifted analysis of
these two detectors's data, as done in O2.
The $\mathcal{F}_{AB}$ estimate is local, using only the last $5$ hr of data,
and therefore tracks possible slow changes in the properties of the detector
noise.
At times when $A$ and $B$ are the only operating detectors, any candidate
events are then submitted directly to GraceDB with false-alarm rate
$\mathcal{F}_{AB}$, the analysis being effectively identical to O2.
However, if additional detectors are observing at the time of the candidate
($C$, $D$\ldots), a combined false-alarm rate is computed as follows.

First, we correct the double-coincident false-alarm rate to include the trials
factor associated with the choice between possible detector pairs,
\begin{equation}
    \mathcal{F}^{\rm tf}_{AB} = {N \choose 2} \mathcal{F}_{AB}
    \label{eq:fartrials}
\end{equation}
where $N$ is the number of detectors that can generate double coincidences at
the time of the candidate.
If multiple instrument pairs generate coincident candidates from the same
transient, we select the candidate having the lowest false-alarm rate. If tied,
we select the candidate with the largest ranking statistic value.
\begin{figure*}[tb!]
    \includegraphics[width=2\columnwidth]{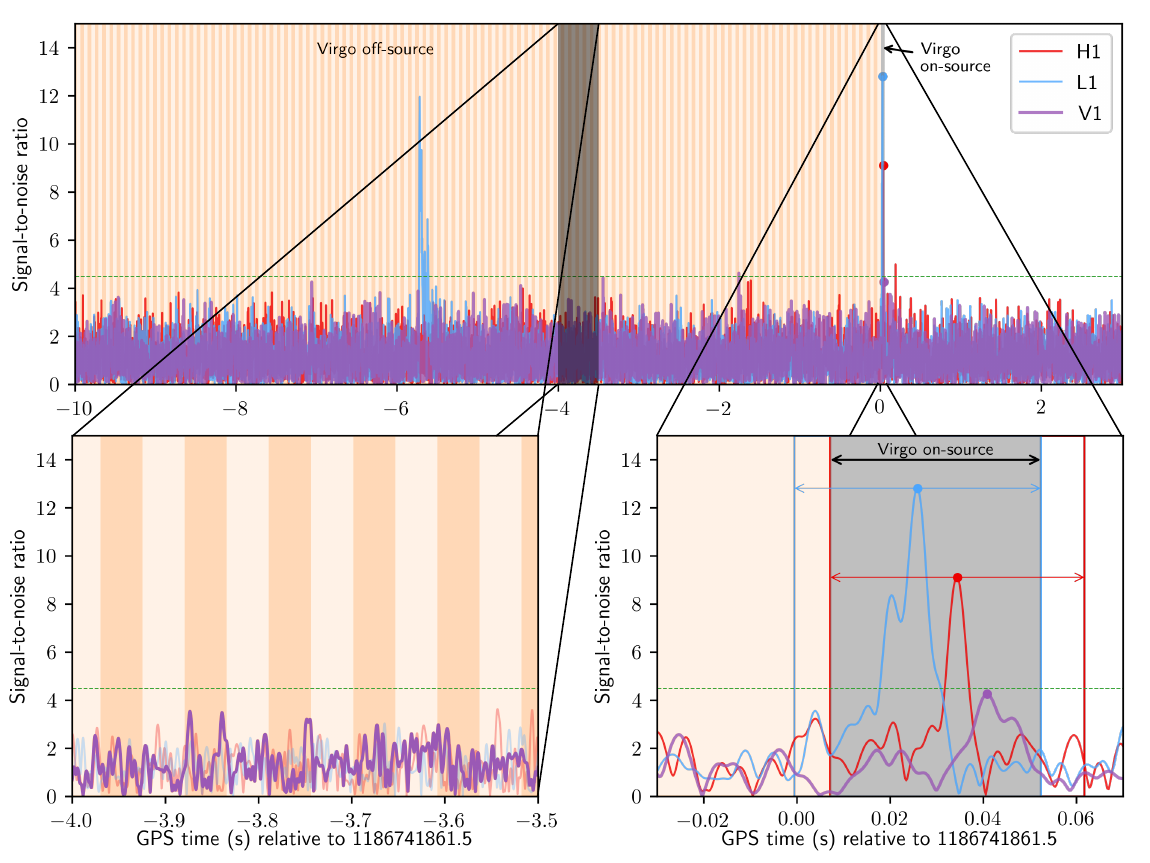}
    \caption{Visualization of the process used by PyCBC Live to generate a
    three-detector coincident candidate.
    In this example, the LIGO-Hanford and LIGO-Livingston SNR time series (red
    and blue curves) have two nearby peaks above the trigger-generation
    threshold (dashed horizontal line). The coincident peaks then produce a
    Hanford-Livingston coincident candidate.
    The window of possible signal arrival times at Virgo (on-source region, dark
    gray in the bottom-right panel) is calculated using the Hanford and
    Livingston triggers, as indicated by the horizontal arrows.
    The Virgo SNR time series is calculated (violet, thick curve) and searched
    for its maximum within the on-source region. Once the maximum is
    identified, its statistical significance is determined by comparing it to
    the maxima occurring in the off-source intervals (vertical stripes
    of alternating color in the top and bottom-left panels).}
    \label{fig:followuppvalue}
\end{figure*}
Then, using the template of the selected candidate, we calculate
the signal-to-noise ratio (SNR) time series for the remaining detectors, which
we call \emph{followup} detectors, and then use this time series to obtain a
p-value for each followup detector.
Assuming a plane GW traveling at the speed of light, and using the arrival times
estimated at detectors $A$ and $B$, an \emph{on-source} time interval of
possible arrival times is determined at each followup detector.
The maximum SNR within the on-source interval is identified.
Next, $150$\,s of data immediately before the on-source window, called
\emph{off-source} data, are segmented into $N_{\rm off}$ intervals of the same
duration as the on-source data (Figure~\ref{fig:followuppvalue}). The maximum
SNR in each off-source interval is calculated, and the number $M_{\rm off}$ of
off-source intervals having a SNR larger than the on-source SNR is used to
compute the p-value,
\begin{equation}
    p_C = \frac{1 + M_{{\rm off,}C}}{1 + N_{\rm off}},
    \label{eq:followuppvalue}
\end{equation}
where $C$ labels the followup detector(s).
This is the probability of producing a SNR as large as the one observed under
the assumption that detector $C$'s data contains only noise.
Such a p-value is produced for each followup detector ($p_C$, $p_D$\ldots).
In addition, we obtain the p-value for the original double-coincident candidate as
\begin{equation}
    p_{AB} = 1 - \exp \left[ -T_{\rm live} \mathcal{F}^{\rm tf}_{AB} \right]
\end{equation}
where $T_{\rm live} = 4.38$\,hr (0.005\,yr) is a fiducial livetime used to convert
false-alarm rates to p-values. Its value is close to the amount of single-detector
data stored for background computation, and also close to the minimum inverse
false-alarm rate required for uploading a candidate to GraceDB. The two-detector
p-value $p_{AB}$ is then combined with the followup detector p-value
$p_C$ using an adaptation of Fisher's method \citep{FisherBook}:
\begin{eqnarray}
    p_{ABC} &=& \mathcal{P} \left( 2, -\ln(p_{AB} p_{C}) \right) = \\
            &=& p_{AB} p_{C} \ \left[ 1 - \ln(p_{AB} p_{C}) \right]
\end{eqnarray}
where $\mathcal{P}(\cdot,\cdot)$ is the regularized gamma function.
(For more than one followup detector, our algorithm performs this combination
iteratively.)
Next, we convert back to a false-alarm rate,
\begin{equation}
    \mathcal{F}_{ABC} = -T^{-1}_{\rm live} \ln(1 - p_{ABC}).
\end{equation}
For the final significance, we additionally select the minimum of the original
two-detector and multi-detector false-alarm rates,
\begin{equation}
    \mathcal{F} = 2 \min \{ \mathcal{F}^{\rm tf}_{AB}, \mathcal{F}_{ABC} \},
\end{equation}
where the trials factor of 2 accounts for this additional choice. This procedure
produces a self-consistent rate of false alarms under the null hypothesis, and
assuming time-invariance of the noise for the collected background. The latter
assumption may be violated if one or more detectors rapidly change to a new
operating state, hence the desire is to use as short a background collection
time as possible to track these changes.

We now apply the above equations to two limiting examples to illustrate possible
results in practice.
We first consider the case of a nearby source and $3$ detectors with similar
sensitivities.
The source produces a very loud coincidence in detectors $A$ and $B$ as well as
a very high SNR in detector $C$, where by ``very loud'' we mean that the SNR is
higher than any background it is compared to.
In this case we obtain the lower bound $(\mathcal{F}_{AB})^{-1} > 100$\,yr,
determined by the amount of data chosen to estimate the double-coincident background.
Assuming an on-source window of $40$\,ms for detector $C$, we also have the
upper bound $p_C \lesssim 2.7 \times 10^{-4}$.
Then, using the above formalism, we obtain $\mathcal{F}^{-1} \gtrsim 3500$\,yr.
Consider now a similar situation, but the sensitivity of detector $C$ is so low
that the signal is completely undetectable in its data, leading to $p_C \approx
0.5$.
This results in the bound $\mathcal{F}^{-1} \gtrsim 17$\,yr, weaker than the
original bound on $\mathcal{F}_{AB}$ due to the combined effect of the trials
factors and a detector with low sensitivity.
Nevertheless, the limit is still well below the threshold required for a public
alert and for considering the candidate worthwile of additional astronomical
observations.
In these examples, we emphasized that the resulting false-alarm rates are to be
understood as upper limits, as they are limited by the amound of data chosen to
estimate the background, and not by the SNR of the candidate.
The limits can be lowered by using more data, but this comes at the risk
of increasing our sensitivity to sudden changes in the noise properties of the
detectors.

Note that the on-source SNR in the followup detectors is not required to cross
any threshold. Hence, this method naturally allows even weak signals in the
followup detectors to increase the significance of the original coincident
candidates.
However, a potential limitation of the method is represented by the
implicit assumption that all detector pairs are equally likely to produce an
initial double coincidence in response to a signal. This leads to the trials
factor in Eq.~\ref{eq:fartrials} being potentially very large: for a network of
3 detectors, such as during O3, the total trials factor can be as high as 6.
As additional observatories join the network, the trials factor will grow rapidly
and this method may not offer a sensitivity close to optimal. One way to reduce
the trials factor would be to use the local sensitivity of each instrument to 
predict the instrument pair(s) most likely to produce a detection, and only
consider those pair(s) in the calculation of the combined false-alarm rate.
An alternative strategy for a heterogeneous network could be modifying 
Eq.~\ref{eq:fartrials} to weight each detector pair in a way that accounts for
different sensitivities. 
Finally, the iterative application of Fisher's method described above could be
replaced by a single application of the method to all available p-values.
Although the two approaches produce p-values well within a factor of 2 in the
case of 2 followup detectors (i.e.~a LIGO-Virgo-KAGRA network), a single
combination may lead to a higher sensitivity in the case of a 5-detector
network.

\subsection{Identification of signals in a single detector}
\label{sec:singles}

During an observation run, there will undoubtedly be times during which a
detector cannot operate, or is affected by severe nonstationary or transient
noise.
Moreover, all detectors are blind to signals coming from certain
positions in the sky, and these positions are detector-dependent.
Therefore, a non-negligible fraction of signals are unable to produce
coincident triggers in two or more detectors
\citep{Callister:2017urp,Nitz:2020naa}.
By chance, two events that were very promising targets for EM followup
observation were first identified as single-detector triggers:
GW170817, initially seen as a single-detector trigger in
LIGO Hanford due to a large glitch affecting LIGO Livingston
\citep{TheLIGOScientific:2017qsa} and GW190425, observed by LIGO Livingston
and Virgo, but too weak to be detectable in Virgo \citep{GW190425}.
In cases like these, we have a single-detector trigger and no coincidence.
We cannot rely on the usual robust time-slide approach to establish the false-alarm
rate of the trigger, and an alternate method is required. We note and caution that
formally, the false alarm rate can only be assigned with confidence to be less than
once per livetime for single-detector candidates. Beyond this point, extrapolation
is used in order to generate low-latency alerts.
If the detector noise changes in unexpected ways, the extrapolation may become
invalid and the rate of false positives for single-detector candidates may no
longer match an extrapolated false alarm rate.

During O2, PyCBC Live relied on a simple algorithm to identify single-detector candidates
based on a pre-determined ranking statistic threshold, and a set of signal-consistency
cuts, which were chosen based on early detector data.
The algorithm was only active when a single detector was observing.
We improve on this method by implementing a complete ranking statistic and
procedure for extrapolating the false-alarm rate. We further
increase the coverage of our single-detector search by allowing it to operate
during times when multiple detectors are observing, as the relative
sensitivities of two detectors may imply that a signal can only be seen in one
of them.

Our single-detector trigger ranking statistic is the usual \emph{reweighted SNR}
\citep{Usman:2015kfa},
\begin{equation}
    \newsnr =
    \left\{
    \begin{array}{lR}
        \rho \left[ (1 + (\chi^2_r)^{p/2})/2 \right]^{-1/p} & if $\chi^2_r > 1$ \\
        \rho & if $\chi^2_r \leq 1$
    \end{array}
    \right.,
    \label{eq:singlerank}
\end{equation}
where $\rho$ is the matched-filter SNR, $\chi_r^2$ is the time-frequency
discriminator described by \citet{Allen:2004gu}, and $p$ is an index set to the usual
value of $6$.
Note that the same $\newsnr$, calculated for each detector and complemented by
other terms, also defines the ranking statistic of coincident triggers.

As we do not have a coincident trigger to corroborate the evidence of a signal,
we must apply strict cuts in order to remove triggers which are likely to be
glitches. Therefore we remove any trigger with $\newsnr < 9$ or $\chi_r^2 > 2$,
which comes at the cost of possibly removing some signals that are not
particularly well matched by the templates.
We further restrict the calculation to triggers coming from a template
describing a system with a non-negligible mass remaining outside of the final
black hole. By doing so, we ignore many of the templates which match best to
common types of glitch in the LIGO detectors, as well as focus on the templates
corresponding to sources which are most interesting for potential generation of
EM counterpart emission.
Applying the remnant mass equation from \citet{Foucart_2018} to all templates in
our bank, we find that this mass is negligible for templates with duration
shorter than $\sim7$\,s.
Therefore, we prevent triggers associated with shorter templates from generating
single-detector candidates.

The probability distribution of $\newsnr$ for triggers associated with detector noise is
expected to follow a falling exponential, as described in
\cite{davies2020extending},
\begin{equation}
    \mathrm{p}(\newsnr | \textrm{noise}) =
    \left\{ \begin{array}{lR}
        \alpha \exp \left[-\alpha (\newsnr - \newsnr_{th})\right] & if $\newsnr > \newsnr_{th}$ \\
        0 & if $\newsnr \leq \newsnr_{th}$
    \end{array}\right.,
    \label{eq:newsnrnoise}
\end{equation}
where $\newsnr_{th}$ is a threshold on the reweighted SNR, and $\alpha$ is the
fit coefficient. Given the selection cuts defined above, we find
empirically that this model describes the detector noise very well.

For each detector, we fit Eq.~\ref{eq:newsnrnoise} to the triggers from a day's
worth of data and record the fit coefficient, combining these over a longer
period of time.
Then, during PyCBC Live's operation, we can use this coefficient to estimate the
false-alarm rate for each single-detector trigger that passes the selection cuts
described above.
In general, this fit could be performed separately for each template, but we
instead choose to perform the fitting in five bins which are spaced
logarithmically in template duration.
This choice allows us to group many templates which behave similarly in the
presence of noise, and hence increase the number of triggers available for the
fit.

Noise in ground-based GW detectors is affected by slowly-varying environmental
processes, like weather, and by commissioning activities that change the
detector properties over time. Therefore, the statistical properties of the
noise are time-dependent, and the fit coefficients for each bin will also vary
over time.  To account for this variation, we combine the daily fit values in
one of two ways. The maximum likelihood choice of $\alpha$ is proportional to
the inverse of the mean $\newsnr$ for each
template~\citep{davies2020extending}, and so we can take the mean of
$\alpha^{-1}$ over the different days, weighted by the number of triggers from
each day in order to combine these fits.  Alternatively, we could take a low
percentile of the $\alpha$ distribution over different days. This would lead to
a much more conservative estimate of the false alarm rate. In our case we use
the $5^\mathrm{th}$ percentile value, and call this the \emph{conservative} fit
coefficient.
This conservative choice would be used for issuing alerts during future observing
runs, as it would reduce the number of false alarms compared to the use of the mean
coefficient.

\begin{figure}[tb!]
    \includegraphics[width=\columnwidth]{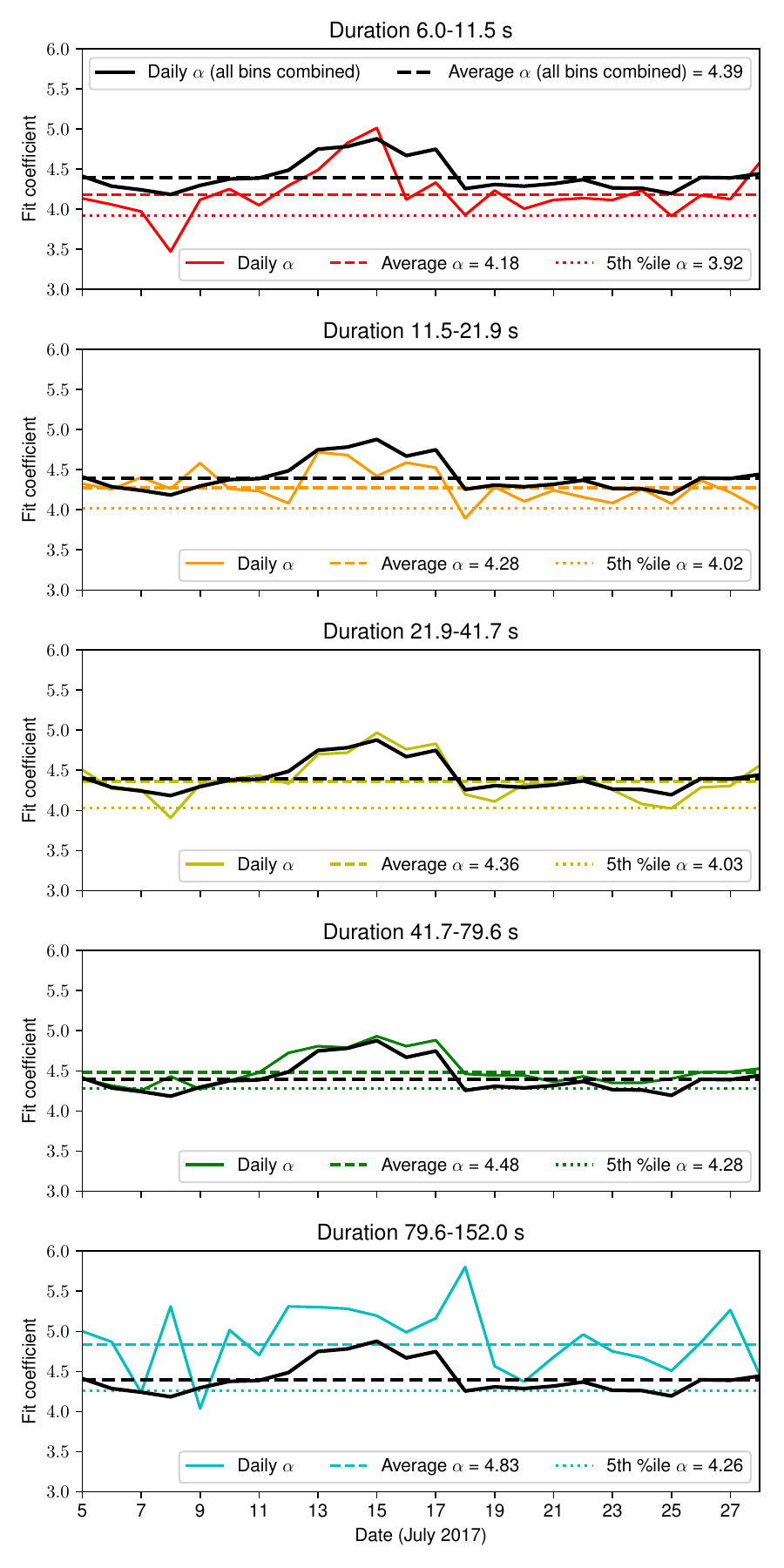}
    \caption{Fit coefficients $\alpha$ calculated daily for
    triggers from a month of O2 LIGO Hanford data
    which meet the $\newsnr$, $\chi^2_r$ and template duration cuts as described in
    the text. One plot is given for each template duration bin,
    and longer templates generally have fewer triggers at high $\newsnr$, so the
    fit coefficients are larger. The dashed and dotted lines show the mean and
    conservative combinations of the fit coefficients, used to estimate the
    false-alarm rates of future single-detector candidates, and the black lines
    are for comparison to the fit values if they were not separated into different
    duration bins.}
    \label{fig:H1SingleFits}
\end{figure}

We see the variation and combination of the trigger fit coefficient in
Figure~\ref{fig:H1SingleFits}, where the fit coefficient in each template
duration bin is plotted for each day of July 2017, along with the mean and
conservative combinations.

To calculate the expected rate of louder triggers, as well as the trigger distribution
of Eq.~\ref{eq:newsnrnoise}, we must estimate the overall rate of triggers. This
is done by simply counting the triggers which pass the specified thresholds in the
daily fits, and for the mean fit, this is simply an addition of the daily trigger
counts. For the conservative fit choice, we choose the $95^\mathrm{th}$ percentile
daily trigger count and multiply by the number of days over which the fit smoothing
is performed.

Using the originally recorded trigger from LIGO Hanford, and the conservative
fit values calculated from the O2 data from July 2017, we estimate the single-detector trigger
false alarm rate in LIGO Hanford of GW170817 to be 1 per $\mathcal{O}(10^9)$\,yr.
If we choose the mean fit combination, then the estimated false alarm rate would
be 1 per $\mathcal{O}(10^{11})$\,yr.

\subsection{Source classification}
\label{ssec:classification}

Source classification between different possible types of coalescing compact
binary is an important element in generating followup alerts for EM or other
counterpart signals.  For the O3 run, four astrophysical source types: BNS, BBH,
NSBH and MassGap, were considered \citep{EMFollowUserGuide}.
The desired classification designated every object with a mass below
$3\,\Msun$ as a neutron star, every object with mass above $5\,\Msun$ as a
black hole, and every object between these two thresholds as a MassGap object;
any binary containing \emph{one or more} MassGap components
was then considered as MassGap.  Accurate classification in low latency is a
considerable challenge: in general for lower-mass binaries, only the chirp mass
\begin{equation}
    \mathcal{M}=\frac{(m_1m_2)^{3/5}}{(m_1+m_2)^{1/5}}
    \label{eq:chirpmass}
\end{equation}
can be precisely measured, while the mass ratio has large uncertainty.
In addition, typically search pipelines report only a point
estimate of redshifted component masses,
and these template masses may be subject to systematic biases relative to the
true source-frame component masses \citep[{see e.g.}][]{Biscoveanu:2019ugx}.
During O3, astrophysical source classification for PyCBC Live candidates was
performed by the LIGO/Virgo rapid alert infrastructure using a
``hard cuts'' method, which assigns Boolean weights (either $1$
or $0$) to the different source types 
based on component mass cuts applied to the reported search template.
The fact that this method neglects uncertainties in component masses and
does not account for any systematic error in mass recovery
%by the search pipeline,
suggests the potential for improvement \citep[compare][]{Kapadia:2019uut}.

A new approach developed during the later part of the O3 run, which will be
described in detail in Villa-Ortega et al.\ (2022, in preparation), uses
the chirp mass recovered by the search pipeline as input, %instead of the component masses of the candidate merger,
and implicitly assumes %as prior information
a uniform density of candidate signals over
the plane of component masses $m_1$, $m_2$. %(between given limits).
These source-frame masses are constrained to the interval $1\,\Msun < m < 45\,\Msun$,
where the lower bound is the lower limit on the template bank mass space described in the
Section \ref{ssec:bank}, and the upper bound is chosen based on BBH population studies up
to the first half of the third observing run, O3a \citep{LIGOScientific:2018jsj,LIGOScientific:2020kqk}.
Constraining the chirp mass to be within a confidence interval around a
point estimate derived from the search template determines an allowed region
in the $m_1$--$m_2$ plane; we then estimate
the probability that a candidate belongs to each source category
to be directly
proportional to the area of the allowed region satisfying the criteria for
the given category.\footnote{Systems with chirp masses outside the considered limits
are assumed to be BNS if they lie below the minimum value, and BBH if they are above the maximum.}
The output of the method for a given event is a list of
probabilities $\left\{P_{\rm BNS},P_{\rm NSBH},P_{\rm MG},P_{\rm BBH}
\right\}$ summing to unity.

To compute these areas we require accurate estimates of the candidate chirp
mass in the source frame, thus we apply a correction for the bias caused by
cosmological redshift.
For this correction we use
estimates of the luminosity distance and its uncertainty derived from the
effective distances \citep{FindChirp} of the trigger(s) comprising the event:
\begin{eqnarray}
    \tilde{D}_{\rm L} &=& \mathrm{const}
        \cdot \min (D_{\rm eff}), \\
    \tilde{\sigma}_{{D}_{\rm L}} &=& \tilde{D}_{\rm L} \cdot \mathrm{const}
        \cdot \rho_{\rm coinc}^{-p},
    \label{eq:lumdistestimation}
\end{eqnarray}
where $\min (D_{\rm eff})$ is the minimum effective distance over all triggers
obtained for a given coincident or single-detector event, and the constants
of proportionality and power law exponent $p$ may be derived from a fit to
previously obtained three-dimensional localizations produced by the BAYESTAR
pipeline \citep{Singer:2015ema} for PyCBC Live events.
Although in most cases the uncertainty in the source chirp mass is expected to
be dominated by the redshift (distance) uncertainty, we combine this with a
nominal, small uncertainty of 1\% in the detector-frame chirp mass relative
to the value recovered by the search pipeline.

Our approach may be compared to the `ellipsoid-based' method outlined in 
\citet{Chatterjee:2019avs}.  The ellipsoidal error region considered there 
accounts for expected uncertainties in the recovered masses and spins in the
limit of Gaussian detector noise and high SNR (though without attempting to
correct for the source redshift).  It was also noted there that 
the actual recovery of parameters other than $\mathcal{M}$ by search pipelines
was subject to significantly larger systematic error, motivating an alternative
machine learning based method.  Our approach is simpler in that we effectively 
consider the uncertainties in such parameters to be infinitely large: we leave
improvements to this approximation to future work.

\subsection{Search for maximum-SNR template}

The rapid spatial localization of GW alerts performed by BAYESTAR
\citep{Singer:2015ema} relies on the knowledge of the template that maximizes
the likelihood assuming stationary Gaussian noise (or equivalently, the
template that maximizes the network matched-filter SNR) for a given candidate.
The template is usually taken directly from the candidate produced by the
low-latency search.
However, when a search like PyCBC Live generates a candidate in response to an
astrophysical signal, it includes both astrophysical priors and the presence of
nonstationary and non-Gaussian noise features into the significance of the
candidate.
Hence, the template immediately associated with a candidate will not necessarily
maximize the network SNR under the assumption of stationary Gaussian noise.
The sparseness of the template bank will generally also drive the reported
template parameters away from the maximum SNR.
This could in principle introduce biases in the rapid spatial localization, and
more generally affect any low-latency result which uses the mass and/or spin
parameters of the search template, such as the source classification we
described above in Section \ref{ssec:classification}.
See \cite{Biscoveanu:2019ugx, Chatterjee:2019avs} for investigations of such
possible biases.

In order to remove some of the sources of these biases, we developed a followup
process that starts after PyCBC Live reports a candidate to GraceDB.
The process reanalyzes a short amount of strain data around the candidate, and
uses differential evolution \citep{DifferentialEvolution} to find the template
parameters that maximize the network SNR.
The maximization explores the mass and spin parameter space in a continuous
fashion, regardless of the placement of the search templates.
Once the optimization converges, or a predefined timeout of $400$\,s is reached
(whichever comes first) a new candidate is uploaded to GraceDB using the best
template found by the optimization.
The new candidate can then be used to generate new spatial localization and
source classification results, free of potential biases from the
initially-reported template.

\section{Evaluating the improved search technique}
\label{sec:results}

In this section we evaluate the impact and performance of the techniques
described in Section \ref{sec:methods} using both simulated and real data.

\subsection{Sensitivity in simulated data}
\label{ssec:sensitivity}
We first characterize the search sensitivity by simulating a
population of astrophysical signals, adding the signals to a portion of
simulated Gaussian noise, analyzing the data with PyCBC Live and counting how
many signals are recovered at a given false-alarm rate.
The noise models correspond to final design sensitivities of Advanced LIGO and
Advanced Virgo \citep{ObsScenario}.
We focus on evaluating the significance calculations described in Sections
\ref{ssec:methods_multi} and \ref{sec:singles}.
To this end, we compare the sensitivity of the search under different network
configurations: HL, HLV, H, L and V, where H, L and V indicate respectively
LIGO-Hanford, LIGO-Livingston and Virgo.
In the HL and HLV configurations, all observatories are assumed to be observing
at the same time.

We construct a population of BNS signals with component masses distributed
uniformly between $1.35\,\Msun$ and $1.45\,\Msun$.
Spins are assumed to be aligned with the orbital angular momenta, and spin
magnitudes are distributed uniformly between $0$ and $0.05$.
The signals are simulated using a waveform model based on post-Newtonian theory.
The sources have isotropic orientation and sky location.
In order to increase the number of detected signals, we distribute the sources
uniformly in \emph{chirp distance} \citep{FindChirp} up to a maximum value.
When computing the sensitive volume, we then weight each source such that the
effective population has a uniform spatial distribution, as described in
\cite{Usman:2015kfa}.

\begin{figure}
    \includegraphics[width=\columnwidth]{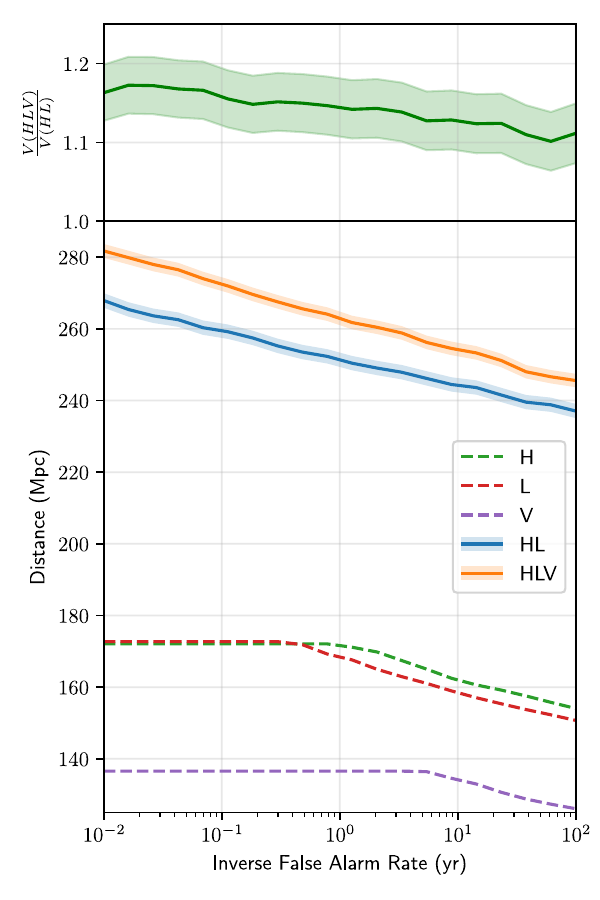}
    \caption{Sensitivity of PyCBC Live, with the O3 configuration, for a
        population of simulated BNS signals added to simulated Gaussian noise
        at design sensitivity. The ``HL'' configuration corresponds to a
        detector network formed by LIGO-Hanford and LIGO-Livingston only.
        The ``HLV'' configuration includes Virgo. The top panel shows the
        relative search volume of the HLV and HL configurations. The bottom
        panel shows sensitivity distances for the multidetector coincidence in
        the HL and HLV configurations (solid lines), and for the single-detector
        triggering (dashed lines). The lighter bands represent the $1 \sigma$
        uncertainties from the Monte Carlo sampling.}
    \label{fig:sensitivity}
\end{figure}

The result of the simulation is shown in Figure~\ref{fig:sensitivity} in terms
of sensitivity distance as well as relative search volume between the HLV and HL
configurations.
We can see that adding Virgo to the LIGO network increases the detection rate of
BNS systems by a few tens of percent under ideal noise conditions.
The single-detector distances are approximately half of what is achieved by a
multidetector network.

\subsection{Sensitivity in real data}
\label{ssec:sensitivity_realdata}

Here we repeat a similar test as presented in Subsection
\ref{ssec:sensitivity}, with the difference that we consider broader ranges of
masses and spins (therefore effectively including BBH and NSBH systems) and we
add their signals to real data from the third observing run of Advanced LIGO
and Advanced Virgo, as opposed to simulated stationary Gaussian detector noise.
We use $\sim 8$ days of O3 data starting from 2019-05-04 13:15:32 UTC.  In the
simulated binaries, neutron stars have masses distributed uniformly between
$1\,\Msun$ and $3\,\Msun$, and spin magnitudes distributed uniformly between 0
and 0.05.  Black holes have masses distributed uniformly between
$3\,\Msun$ and $100\,\Msun$, and spin magnitudes distributed uniformly between
0 and 0.985.  Spins are aligned with the orbital angular momenta in all cases.
BNS signals are simulated using a post-Newtonian waveform model, while NSBH and
BBH signals use the SEOBNRv4\_opt inspiral-merger-ringdown
model~\citep{Devine:2016ovp,PhysRevD.95.044028}.

\begin{figure}
    \includegraphics[width=\columnwidth]{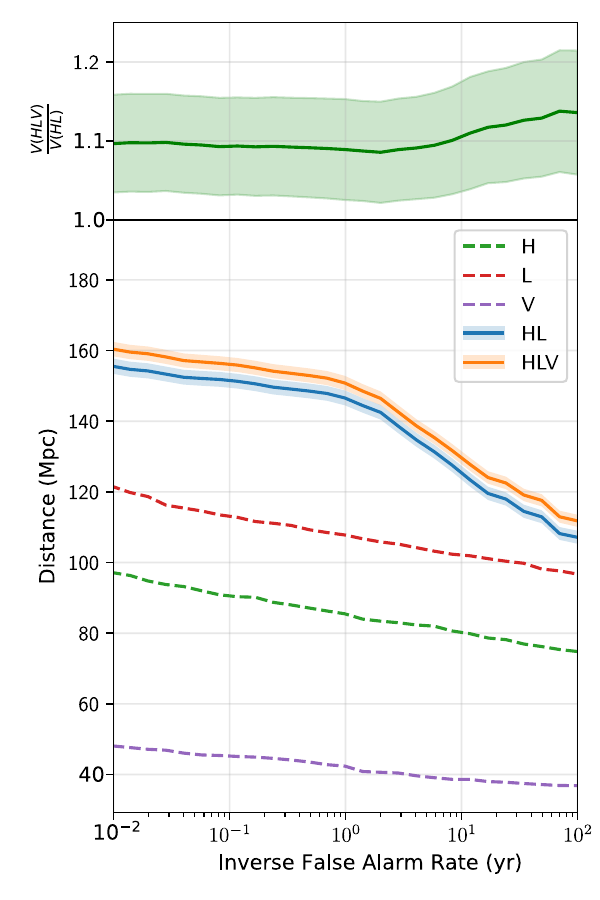}
    \caption{Comparison of PyCBC Live's sensitivity to BNS mergers using an
    O2-like (``HL'') and the final-O3 (``HLV'') search configurations.
    The simulated signals are added to real data from Advanced LIGO and
    Advanced Virgo's third run. The lighter bands represent the $1 \sigma$
    uncertainties from the Monte Carlo sampling.}
    \label{fig:sensitivityO3a}
\end{figure}

The resulting sensitivity for BNS mergers is shown in
Figure~\ref{fig:sensitivityO3a}.
We can see that the improvement in detection rate is around 10\% at false-alarm
rate thresholds relevant for public alerts. This estimate is consistent with
the earlier estimate from simulated data at design sensitivities. NSBH and BBH
mergers, albeit arguably less interesting for rapid alerts, show similar
relative improvements at the same false-alarm rate threshold, and are not shown
here.

\subsection{Redetection of GW170814 and GW170817}

PyCBC Live in the O3 configuration detects GW170814 in O2 data as a coincidence
between the two LIGO detectors, with a LIGO-Virgo network SNR of $\approx 15$.
With the method described in Section \ref{ssec:methods_multi}, and using 5 hours
of lookback background, the candidate is assigned a false-alarm rate of 1 in
$\approx 200$\,yr.

The case of GW170817 is more interesting because of the glitch affecting the
LIGO-Livingston data seconds before merger \citep{TheLIGOScientific:2017qsa}.
Although the glitch is automatically gated by PyCBC Live, the surrounding data
are still flagged as affected by a glitch by the low-latency data-quality flags,
preventing a LIGO double coincidence from taking place.
In addition, the small Virgo SNR also prevents a double coincidence between
LIGO-Hanford and Virgo.
Nevertheless, using the method described in Section \ref{sec:singles}, GW170817
is reported as a LIGO-Hanford single-detector candidate with false-alarm rate of
1 in $\approx 10^9$\,yr.

PyCBC Live can also be configured to ignore data-quality flags.
Under this configuration, GW170817 is detected instead as a LIGO double
coincidence and is assigned a false-alarm rate lower than $1$ in
$\approx 17$\,yr by the method of Section \ref{ssec:methods_multi}.
The much higher value with respect to the single-detector candidate is due to
the upper limit on the double-coincident false-alarm rate imposed by the
duration of the lookback background ($1$ in $100$\,yr), combined with a trials
factor of 6 caused by having 3 observing detectors at the time of the candidate.
The absence of a detectable signal in Virgo produces a relatively large followup
p-value (see Eq.~\ref{eq:followuppvalue}), which cannot overcome the
penalty of the trials factor.
In fact, this situation matches our second numerical example in Section
\ref{ssec:methods_multi}.
Hence, when comparing these GW170817 false-alarm rates, one has to bear in mind
that the LIGO-Hanford-only rate is an extrapolation from months of data, while
the Hanford-Livingston-Virgo rate is an upper limit based on just $5$\,hr of data.

In both configurations, however, GW170817 is reported with a false-alarm rate
well beyond what is required to issue a public alert on the GCN and to consider
the candidate worthwhile of followup observations.

\subsection{Latency}

We measure the latency of the analysis by repeatedly replaying a week of O2
data and analyzing it with PyCBC Live, thus simulating an actual observing run
with a realistic computing configuration.
The test amounts to a total wall-clock time of approximately $50$ days.
For each candidate uploaded to GraceDB, we calculate the latency as the
difference between the upload time recorded by GraceDB, and the merger time
estimated by PyCBC Live.
This quantity includes the processing time in PyCBC Live, as well as the latency
due to the generation and distribution of the replayed O2 data.
We expect the former to be the dominant contribution.

\begin{figure}
    \includegraphics[width=\columnwidth]{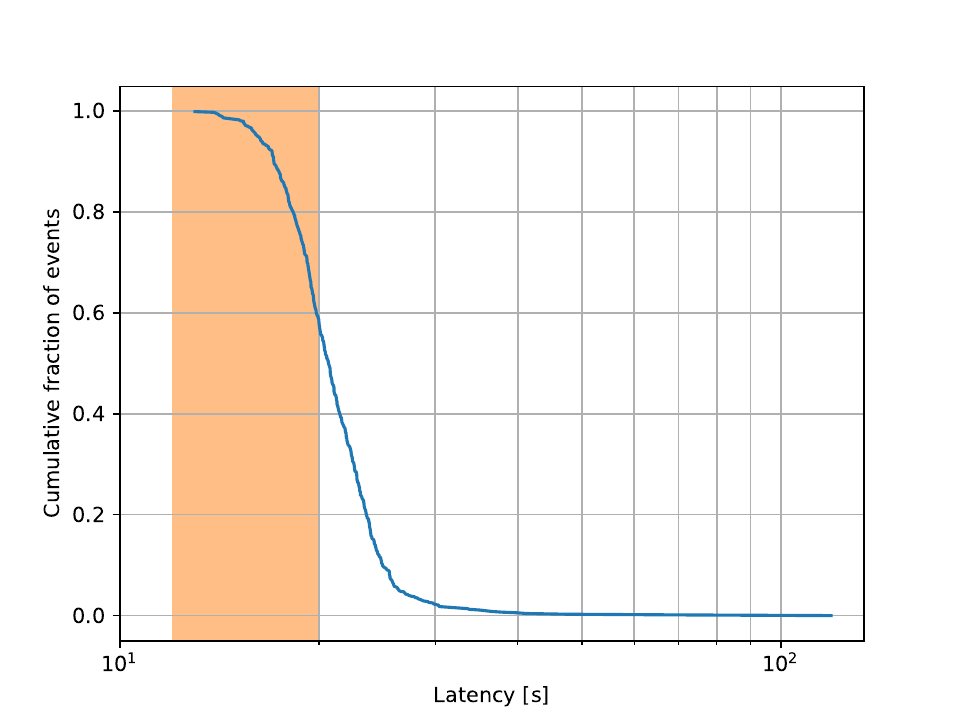}
    \caption{Cumulative distribution of PyCBC Live's latency, for a period
        of replayed O2 data analyzed using the O3 configuration. The latency
        is defined here as the time elapsed between the estimated coalescence
        time of a candidate and the time of creation of the corresponding
        GraceDB event. The shaded region is the range of expected latency due
        to PyCBC Live alone, as described in \cite{Nitz:2018rgo}.}
    \label{fig:latency}
\end{figure}

The cumulative latency distribution is shown in Figure~\ref{fig:latency}.
Most candidates are available in GraceDB within a few tens of seconds after
coalescence, as expected.
The tail extending to $\approx 100$\,s is due to occasional and temporary issues
with the computing infrastructure running the test, typically starvation of the
available computational resources or interruptions of network connections.
A similar tail is also found in a typical production analysis.

\subsection{Astrophysical classification}

\begin{figure*}
    \begin{center}
        \includegraphics[width=0.49\textwidth]{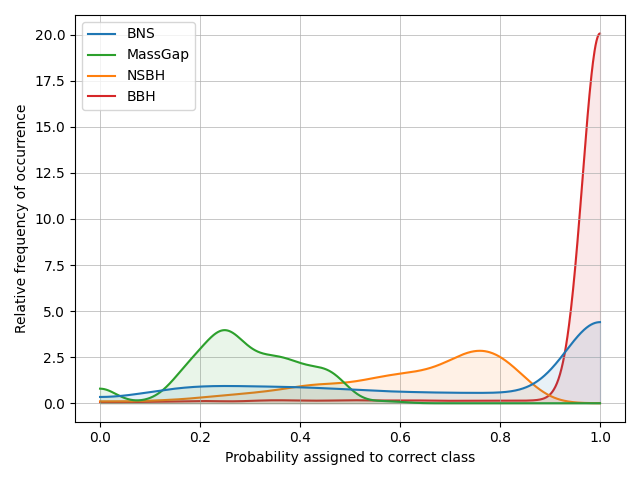}
        \includegraphics[width=0.49\textwidth]{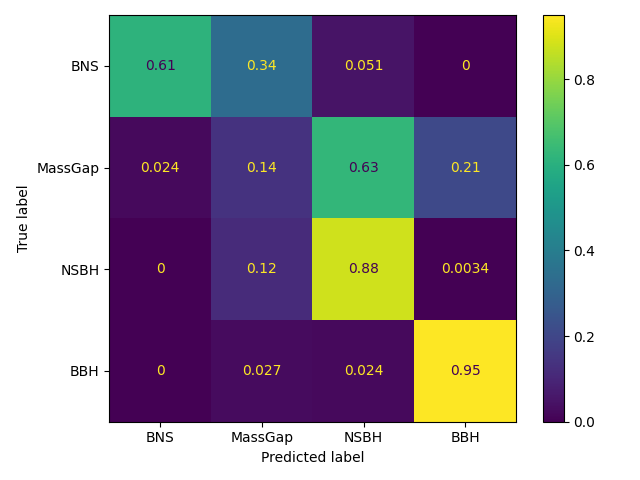}
    \end{center}
    \caption{\textit{Left:} For simulated signals of each source type,
    we show the probability density distribution of assigned probabilities
    for the recovered event to be of the same (i.e.\ correct) type.
    \textit{Right:} Confusion matrix for the source classification assigned
    the highest probability for each simulated signal.
    \label{fig:class_plots}}
\end{figure*}

In order to test the chirp-mass-based classification of Section~\ref{ssec:classification}
we used the same population of simulated signals described in
Section~\ref{ssec:sensitivity_realdata}, with an additional constraint.
Given the $1\,\Msun < m < 45\,\Msun$ limits imposed on component masses, 
the results for asymmetric high-mass NSBH systems outside these limits are not
representative of its accuracy of the method: here, we restrict the black
hole components of simulated NSBH events to be below $50\,\Msun$.
For each simulation recovered by the search, we find the estimated source
probabilities $\left\{P_{\rm BNS},P_{\rm NSBH},P_{\rm MG},P_{\rm BBH}
\right\}$.
We then consider the figures of merit shown in Figure~\ref{fig:class_plots}.

The first figure of merit is the distribution of probabilities for correct
classification, plotted for each category as determined by the true source
masses.
The great majority of BNS and BBH simulations are assigned high or very high
correct class probabilities, as expected given the positions of the target
classes in the $m_1$--$m_2$ plane.  No NSBH simulations are assigned very high
probabilities of NSBH origin, since their contours of constant chirp mass
always overlap other source target regions to some extent, but the majority
are assigned $P_{\rm NSBH}>0.5$.  In contrast, the majority of MassGap simulations
are assigned $P_{\rm MG}\lesssim 0.5$; this again is expected due to the narrow
extent of the MassGap region and its very high overlap with other source target
regions.

The second figure of merit is the correctness of the highest estimated
probability for each simulation, i.e.\ the most likely source category as
assigned by our method.  Comparing this highest probability to the true (target)
classification determined by the simulated source masses, we construct the
confusion matrix shown in the right panel of Figure~\ref{fig:class_plots}.
Simulations of all
source types except MassGap are assigned most likely classifications that are correct
in a large majority of cases; MassGap simulations are, though, preferentially
assigned as most likely to be NSBH.  Given the very high uncertainties
on the rates and masses of actual NSBH and MassGap sources (e.g.\ \citealt{GW190814}),
this bias can be argued to be acceptable, and will also yield a conservative
outcome as the method will err on the side of recommending EM followup for
signals consistent with NSBH origin even if the true source class is MassGap.

\section{Discussion}
\label{sec:discussion}

We described how the PyCBC Live analysis was improved with respect to the O2
configuration between the end of O2 and the end of O3.
The most significant changes are the inclusion of more than two detectors in the
significance calculation, which allowed Virgo to play a prominent role in
the generation of O3 candidates, and the ability to assign significances based
on data from a single detector.
The single-detector significance calculation and source classification methods
were not used during O3 due to its premature end, but they are ready to be
utilised in future runs.
We evaluated these improvements in multiple ways: first by recovering simulated
signals added into ideal detector noise, then by recovering simulated signals
added into real O3 data, and finally by reanalyzing a segment of O2 data
containing the GW170814 and GW170817 events, and discussing how these events
are detected by the new analysis.
The improvements we introduced do not impact the latency.

During O3, $56$ alerts were issued on the GCN without retraction.
PyCBC Live contributed to $34$ of them.
Only $1$ of the $24$ retracted alerts was produced by PyCBC Live.

The next observing run will include a brand new detector, KAGRA.
Unless Virgo and KAGRA reach a sensitivity comparable to LIGO, the larger
resulting detector network will probably warrant further development of the
multidetector significance calculation in order to limit the impact of trials
factors, as discussed earlier.

In preparation for the next observing run, we plan to investigate further
improvements in the handling of instrumental transients.
In particular, the inpainting method presented by \cite{Zackay:2019kkv} could
potentially broaden the applicability of gating to a wider class of glitches, if
it is compatible with the latency requirements of PyCBC Live.
We also plan to study the impact of applying data-quality flags to the online
analysis, and to characterize the effect of the SNR maximization implemented
during O3.

Our proposed rapid source classification method could also be extended to
consider template parameters other than the chirp mass, although these are
subject to higher statistical and systematic errors \citep[{e.g.}][]{Biscoveanu:2019ugx}.
A possible implementation using a larger set of triggers associated with a
given astrophysical event to quantify parameter errors was shown in
\cite{Stachie:2021noh}.

Finally, as the latency of the entire LIGO-Virgo public alert system keeps being
reduced, further development is also under way to reduce the latency of PyCBC
Live.
In particular, the so-called ``early-warning'' detection of BNS mergers
\citep{Cannon:2011vi, Sachdev:2020lfd, Magee:2021xdx} has been implemented in
PyCBC Live as well \citep{Nitz:2020vym}, and will be optimized and
characterized in a future study.

\begin{acknowledgments}
We are grateful to Stuart Anderson, Juan Barayoga, Patrick Brockill and Sharon
Brunett for computing assistance during the operation of PyCBC Live in O3.
We also thank Gregory Mendell and Alan Weinstein for reviewing PyCBC Live's O3
configuration, and Francesco Pannarale for comments on the manuscript.

TDC was supported by an appointment to the NASA Postdoctoral Program at the
Goddard Space Flight Center, administered by Universities Space Research
Association under contract with NASA, during part of this work.
GSCD, VVO and TD acknowledge financial support from the Maria de Maeztu Units of 
Excellence program MDM-2016-0692, from Xunta de Galicia (Centro singular de 
investigación de Galicia accreditation 2019-2022), and from European Union ERDF. 
GSCD and IWH acknowledge the STFC for funding through grant ST/T000333/1.

We acknowledge the Max Planck Gesellschaft and the Atlas cluster computing team
at AEI Hannover for computing support in preparation of this draft.
We are also grateful for computational resources provided by the LIGO Laboratory
and supported by National Science Foundation Grants PHY-0757058 and PHY-0823459.

This paper has LIGO document number LIGO-P2000296 and Virgo document number
VIR-0693E-20.
\end{acknowledgments}

\bibliography{references}

\end{document}